\documentclass[aps,twocolumn,prl,floatfix,superscriptaddress,showpacs,longbilbiography]{revtex4-1}
\usepackage[utf8]{inputenc} 
\usepackage{amssymb}

\usepackage{graphicx}
\usepackage{dcolumn}
\usepackage{braket}
\usepackage{bm}
\usepackage{amsfonts}
\usepackage{amsmath}
\usepackage{amssymb}
\usepackage{color,soul}
\usepackage{wasysym}
\usepackage{mathrsfs}
\usepackage{times}
\usepackage[dvipsnames,svgnames,table]{xcolor}
\usepackage{ulem}
\usepackage{hyperref}
\usepackage{fancyhdr}
\usepackage{float}
\usepackage[english]{babel}
\usepackage{hyperref}
\usepackage[caption=false]{subfig}
\usepackage{soul}

\usepackage{wrapfig}
\definecolor{light-gray}{gray}{0.95}
\usepackage{tcolorbox}

\hypersetup{
pdfstartview={FitH},
pdftitle={2017},    
pdfauthor={LEFFT},  
colorlinks=true,    
linkcolor=NavyBlue, 
citecolor=Maroon,   
filecolor=NavyBlue, 
urlcolor=NavyBlue   
}

\newcommand{\bea}{\begin{eqnarray}}
\newcommand{\eea}{\end{eqnarray}}

\usepackage[framemethod=tikz]{mdframed}

\colorlet{mycolor}{Yellow!60!} 

\begin{document}


\title{Directional control of charge and valley currents in a graphene-based device}

\author{M. Berdakin}
\email{matiasberdakin@gmail.com}
\affiliation{Departamento de F\'{\i}sica, Facultad de Ciencias F\'{\i}sicas y Matem\'aticas, Universidad de Chile, Santiago, Chile}

\author{J. E. Barrios Vargas}
\affiliation{Departamento de F\'{\i}sica, Facultad de Ciencias F\'{\i}sicas y Matem\'aticas, Universidad de Chile, Santiago, Chile}
\author{L. E. F. Foa Torres}
\email{luis.foatorres@uchile.cl}
\affiliation{Departamento de F\'{\i}sica, Facultad de Ciencias F\'{\i}sicas y Matem\'aticas, Universidad de Chile, Santiago, Chile}

\begin{abstract}

\begin{mdframed}[hidealllines=true,backgroundcolor=mycolor!20]
\vspace{0.4cm}
We propose a directional switching effect in a metallic device. To such end we exploit a graphene-based device with a three-terminal geometry in the presence of a magnetic field. We show that unidirectional charge and valley currents can be controlled by the Fermi energy and the magnetic field direction in the active device. Interestingly, unidirectional transport of charge and valley is generated between two-terminals at the same bias voltage. Furthermore, we quantify the valley depolarization as a function of disorder concentration. Our results open a way for active graphene-based valleytronics devices.  
\vspace{0.4cm}

\end{mdframed}


\end{abstract}

\date{\today}
\maketitle


\section{Introduction}
Most often, charge transport in an electronic device is reciprocal, this is, the transmission probability from left to right equals the one from right to left. This arises typically because a device described by a $2 \times 2$ time-independent scattering matrix always satisfies reciprocity~\cite{Buettiker1988}. However, studies in optics~\cite{Leonhardt2006,Pendry2006} and acoustics~\cite{Cummer2007} have broken reciprocity allowing for unidirectional transport. Interestingly, the potentials originating this effect might also remain invisible to an observer analyzing the reflection on one side of the device, a phenomenon termed unidirectional invisibility~\cite{Lin2011}. This has flourished more recently in the context of chiral nanophotonic waveguides~\cite{Petersen2014,Price2018}.
In electronics, in contrast, only a few studies~\cite{FoaTorres2016,DalLago2017} have pointed out ways of emulating this effect and the potential of such one-way devices for the transport of charge, spin and even energy remains mostly unexplored.

Moreover, in materials like graphene, the electronic states close to the Fermi energy belong to one of two inequivalent pockets in momentum space called \textit{valleys}~\cite{CastroNeto2009,Dubois2009,FoaTorres2014}. This binary spin-like degree of freedom is a center of interest. Studies span a plethora of topics, from topological valley effects~\cite{Xiao2007,Yao2008,Gorbachev2014}, to valley selective circular dichroism~\cite{Cao2012}.
\textit{Valleytronics}~\cite{Schaibley2016,cresti_charge_2016} is a promising field though there are still many challenges in both generating~\cite{Rycerz2007a,Gunlycke2011,Ingaramo2016} and detecting valley polarized currents~\cite{Gorbachev2014,Shimazaki2015,Li2016,Komatsu2018,marmolejo-tejada_origin_2017,Wu_intrinsic_2018}.

Here, we propose an \textit{active} graphene-based device producing unidirectional invisibility in the transport of charge and valley. The proposed device allows to control current flow in a peculiar way: The switch-off mechanism does not rely on the gap of the device material, this is, as a transistor the current can be switched off even when the device is effectively metallic. Furthermore, the output is directional, very much as in a diode though here there is no bias voltage applied between the source and drain.

\begin{figure}
  \centering
  \includegraphics[width=\columnwidth]{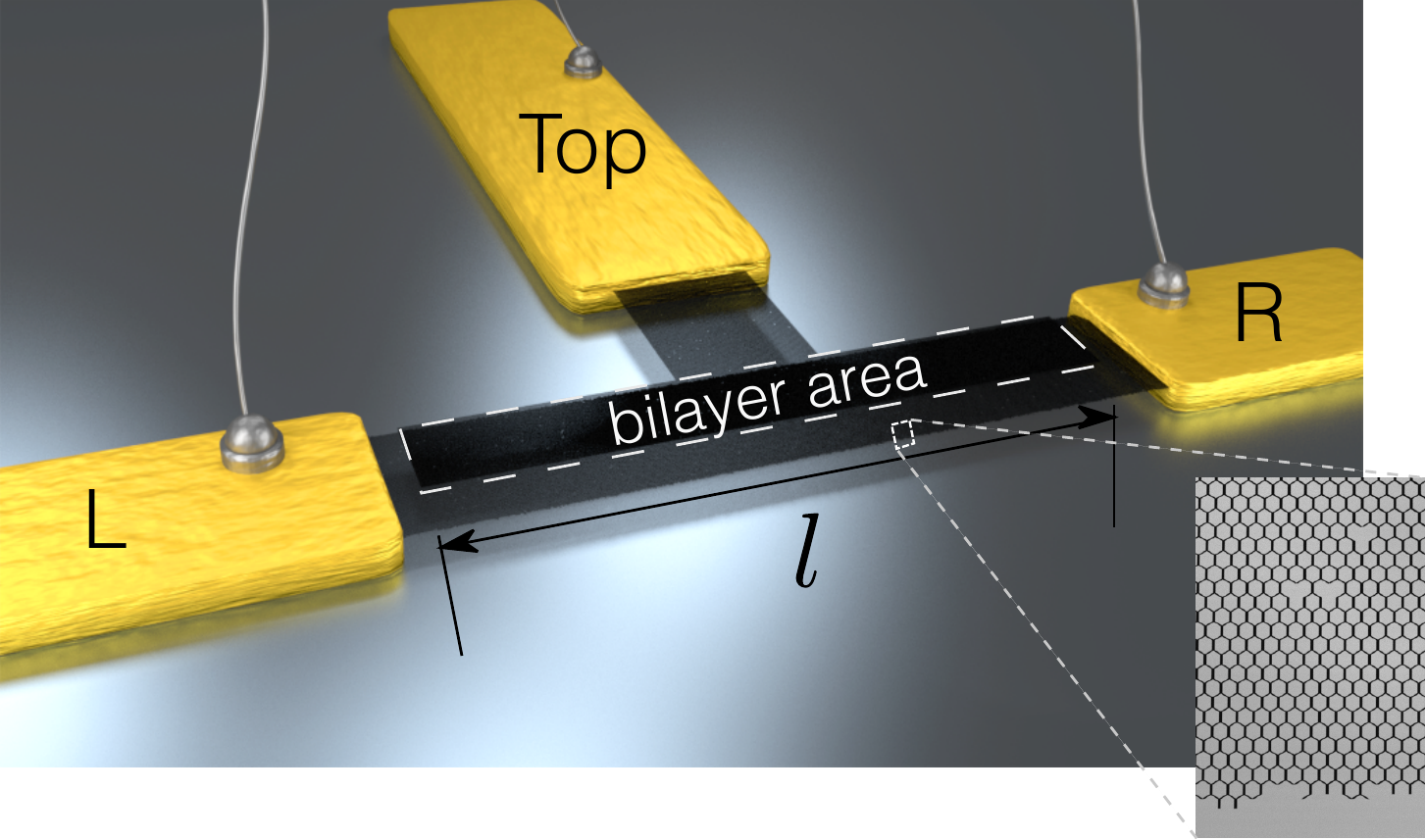}
  \caption{Three-terminal geometry setup, where the
top layer covers just one edges. Furthermore, to probe the
robustness of this setup, the device contains roughness on the bottom layer and random
disorder (0.3$\%$ vacancies to the overall sample), see inset.}
  \label{fig1}
\end{figure}

As in a transistor, our proposal uses three-terminals, that we call left (L), right (R) and top (see figure ~\ref{fig1}). Unlike a conventional field-effect transistor, the top lead generally carries a current as we show later on. In this sense, it may resemble a bipolar junction transistor but in contrast to it, here we do not obtain a current gain, but the ability to turn on or off the current in a given terminal thereby qualifying as an active device. The central part of the device is formed by a monolayer-bilayer junction threaded by a magnetic flux and connected to monolayer graphene leads on the bottom ($L$ and $R$) and top ($T$) layers. In the sense of Ref.~\cite{Basov2017}, we use heterostructuring~\footnote{The fact that the system is a graphene bilayer is accessory, as the setup works also with a metallic 2D material on top of graphene.} and high magnetic fields to achieve \textit{on-demand} properties. A valley polarized current flows unidirectionally from L to R which remain at the same chemical potential. The current and valley can be inverted by tuning the Fermi energy and the magnetic field applied to the device, thereby allowing for a charge and valley switching device working at zero bias voltage. 

In the following we will show how the third lead can be used as a handle to allow for one-way transport among the left and right leads. The bilayer region is used to provide a selective environment for the currents flowing through the covered edge. Essentially, the states on the top layer hybridize with those at the edge of the lower layer, thereby diverting the current from the bottom layer to the top one. This effect, however, would not be effective without the addition of a third terminal (the top one) so that the current is allowed to escape.

Now we introduce in detail our model and the simulation scheme, and our main results. Since the valley currents are expected to be specially sensitive to defects producing intervalley scattering, we also devote a section to study the valley polarization as a function of the concentration of defects in the device. Our study expand the already reach prospects for electronics based on 2D materials~\cite{Fiori2014,Iannaccone2018}.

\begin{figure*}
  \centering
  \includegraphics[width=15cm]{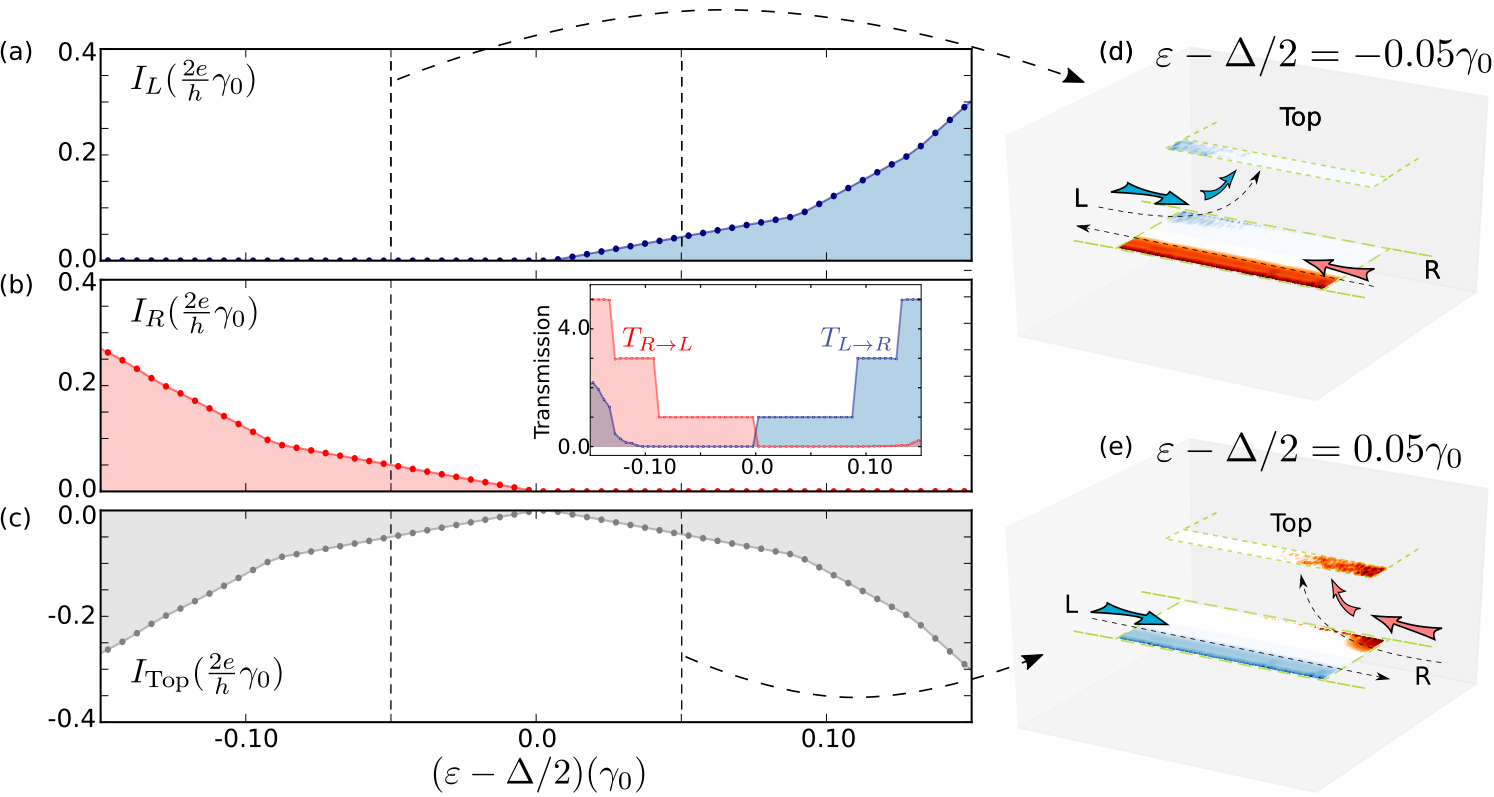}
  \caption{Panels (a), (b) and (c) show the currents in the left, the right, and the top leads as a function of the Fermi energy in the left and right leads $\varepsilon_F$ (set to be equal) measured from $\Delta/2$ (the onsite energy on the lower layer). These results are for a device width $W=104a$ ($\sim 25$ nm) and length $L=512 a$ ($\sim$ 125 nm), the perpendicular bias is set by $\Delta$ of 0.4$\gamma_{0}$ while a magnetic flux of $0.0055\Phi_0$ per plaquette perpendicular to the device is applied. The top layer covers just $W/4$ of the bottom one, edge roughness and $0.3\%$ of vacancies are added. The inset in (b) shows the transmission probabilities as a function of the incident electronic energy $\varepsilon$ minus $\Delta/2$, emphasizing the directional effect. (d) and (e) show modulus squared of the scattering wave functions at two different energies. Each plot shows the scattering wave functions for electrons coming from $L$ and $R$ in two color scales for better visualization (from deep red to white for right incidence and from deep blue to white for left incidence, both in logarithmic scale and spanning five orders of magnitude change). Depending on the energy electrons coming from one direction are either fully transmitted or fully diverted to the top.}
  \label{fig2}
\end{figure*}


\section{Model and method} 

Our simulations are based on a tight-binding model for a device as the one represented in Fig.~\ref{fig1}. Specifically, we use a nearest-neighbors Hamiltonian with hopping $\gamma_{0}=2.7$eV~\cite{Dubois2009,FoaTorres2014} described the Hamiltonian:

\begin{equation}
\label{Hamiltonian}
{\cal H}=\sum_{i}E_{i}^{{}}\,c_{i}^{\dagger}c_{i}^{{}}-\sum_{\left\langle i,j\right\rangle} [\gamma_{i,j} c_{i} ^{\dagger}c_{j}^{{}}+h.c.]+{\cal H}_{\perp},
\end{equation}
where $c_{i}^{\dagger}$ and $c_{i}^{{}}$ are the electronic operators for creation and annihilation at the $\pi$-orbital on site $i$, $\gamma_{i,j}$ is the nearest-neighbors matrix element between sites $i$ and $j$. ${\cal H}_{\perp}$ models the Bernal stacking with $\gamma_{1}$ being the matrix element between the $A$-type atom on the top layer and the $B$-type atom on the lower one. Additionally, $\gamma_{i,j}$ includes the effect of a magnetic field applied perpendicular to the structure (by means of a Peierls substitution):

\begin{equation}
\gamma_{ij}=
\gamma_{0}\exp\left(\mathrm{i}\frac{2\pi}{\Phi_0}\int_{
\bm{r}_i}^{\bm{r}_j}\bm{A}\cdot\mathrm{d}\bm{\ell}\right)\,,
\label{gama}
\end{equation}
where $\Phi_0$ is the magnetic flux quantum. To avoid reflections the magnetic field is applied to the whole device including the leads. The magnetic field is chosen so that the decay length of the edge states towards the sample $\xi$ is much smaller than the sample width $W$ ($\xi\sim 2-3a$). In this multi-terminal case one needs to smoothly change the gauge so that it preserves the translational invariance on each terminal. To such end we use a Landau gauge following the prescription given in Ref.~\onlinecite{Baranger1989}.

To model the bias voltage applied perpendicularly to the bilayer the on-site energies $E_{i}$ are modified, being $\Delta/2$ for the lower layer and $-\Delta/2$ for the upper one (the same bias is applied to the corresponding leads). 

The transport simulations correspond to the coherent regime, where Landauer-B\"uttiker theory predicts that the current at lead $\beta$ is:

\begin{equation}
\label{current}
{\cal I}_{\beta}=\frac{e}{h} \int \sum_{\alpha \neq \beta}[T_{\alpha \rightarrow \beta}(\varepsilon) f_{\alpha}(\varepsilon)- T_{\beta \rightarrow \alpha}(\varepsilon) f_{\beta}(\varepsilon)] d\varepsilon,
\end{equation}
where $\alpha,\beta=L,R,T$, with $T$ being the top lead, and $f_{\alpha}$ is the Fermi-Dirac distribution at lead $\alpha$.

The results presented here were obtained with home-made codes built on the Kwant~\cite{Groth2014} module. Two of the leads are plugged to the lower layer ($L$ and $R$) and a third one ($Top$) to the upper layer. In our simulations the device width ($W$) and length ($\ell$) are 104$a$ ($\sim$ 25 nm) and 512$a$ ($\sim$ 125 nm), respectively, and the top layer covers just $W/4$ of the bottom one. Furthermore edge roughness and $0.3\%$ site vacancies were added to the sample region as shown in the inset of Fig.~\ref{fig1}. The perpendicular bias between the bottom and top $\Delta$ of 0.4$\gamma_{0}$, and a magnetic field was applied perpendicularly to the device. It is important to note that these parameters were chosen for easier computation and to show the mechanism. Nevertheless, one may drastically reduce the magnetic field and with that the bias voltage to experimentally more accesible values just by using a larger sample. Finally, we would like to mention other related proposal~\cite{Mahfouzi2010} involving a ferromagnet on topological insulator and another where stacking variations are used to achieve layer selectivity in transport~\cite{Abdullah2016}.

\begin{figure*}[t]
  \centering
  \includegraphics[width=12cm]{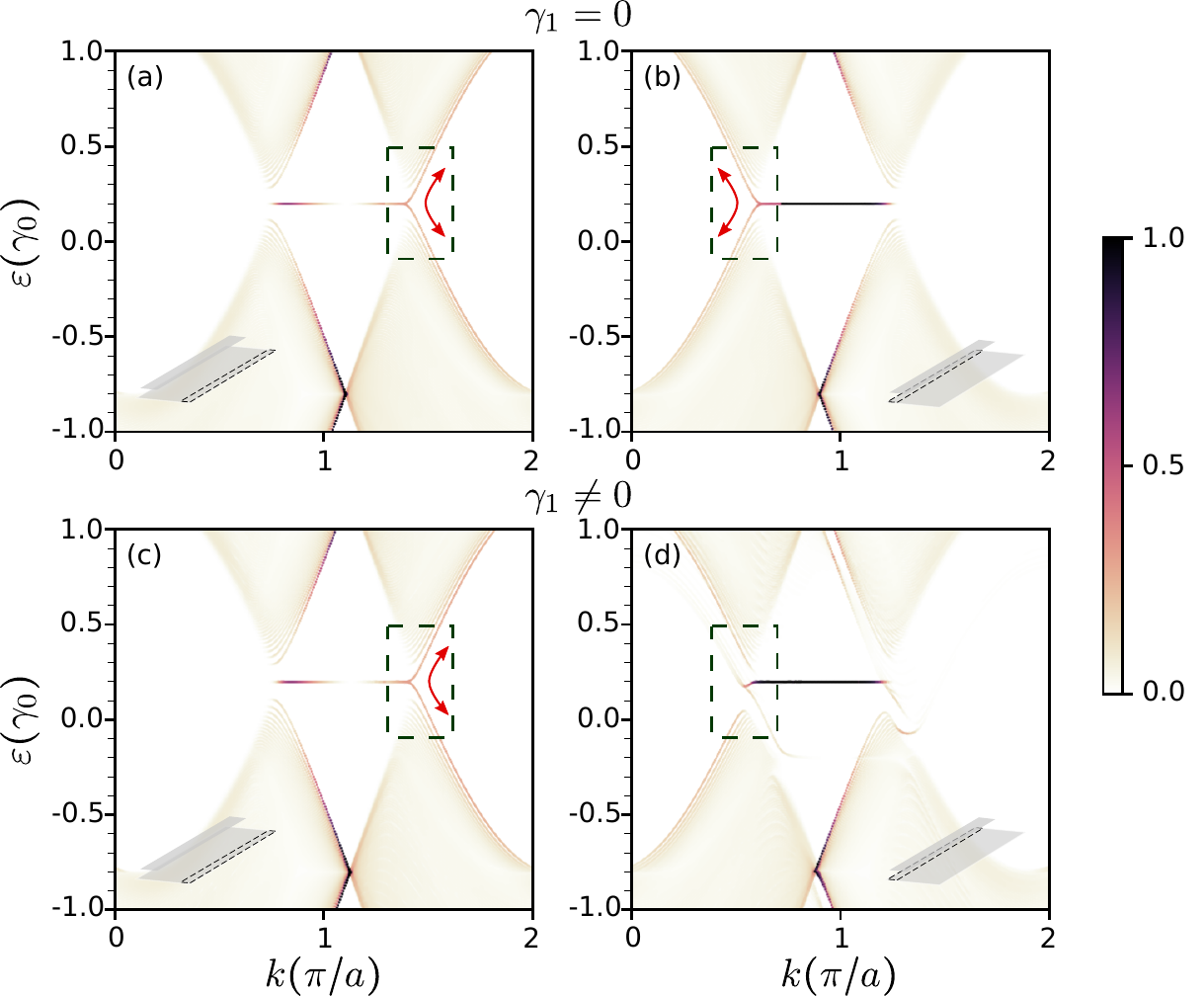}
  \caption{Dispersion of a “zig-zag” graphene bilayer ribbon with a color scale encoding the weight of each state on the area demarked with dashed lines in the insets of each panel. The interlayer interaction is switched off in the a and b panels while is set to $\gamma_{1}$ = $0.15 \gamma_{0}$ in the c and d panels. The width (W) of the bottom layer is 150$a$ and partially covered by a top layer of W/4. The structure is threaded by a magnetic field, while a bias voltage is set perpendicularly to bilayer ($\Delta=0.4\gamma_{0}$). Note the difference between the states in the dashed box with and without the interlayer coupling.}
  \label{fig3}
\end{figure*}


\section{Results and discussion} 

\subsection{Directional transport and non-reciprocal bandstructure} 

Figure~\ref{fig2}(a-c) shows the currents on each lead as a function of the occupations set by the Fermi energy on the left and right leads (measured from $\Delta/2$, the onsite energy at the lower layer). Note that $L$ and $R$ are kept at the same chemical potential while the bias voltage with the top lead is kept at a fixed value. When the Fermi energy is set to $\Delta/2-0.05\gamma_0$ so that the right lead has a non-zero current the left one is inactive (\textit{off}) and viceversa. The origin of this sharp asymmetry can be visualized in Fig.~\ref{fig2}(d) showing the probabilities associated to scattering states coming from the right (left) is shown in red (blue) tones. It is important to note that Fig. 2 (d) and (e) represent the scattering wavefunctions incoming from left and right leads into the bilayer-monolayer junction (the color scale is in log scale for better visualization). The positions of each lead plugged to the scattering zone has been emphasized with L, R and Top, located along the green dashed lines. The chirality of the states induced by the magnetic field separates these scattering wave-functions according to the direction of incidence, while the geometry of the top layer determines their fate: The states incident from the right lead are fully transmitted to the left lead. In contrast, those states incident from the left lead are not transmitted to the right but are neither reflected back to the left (see Fig.~\ref{fig2}(b)-inset showing the transmission probabilities as a function of the incoming electronic energy measured from $\Delta/2$), they are, in turn, transmitted to the top electrode. Thus, transport becomes directional, $I_R$ is nonzero while the lack of reflection in the left lead renders a vanishing $I_L$. The opposite occurs when $\varepsilon_F=\Delta/2+0.05\gamma_0$ as seen in Fig.~\ref{fig2}(e). Note that, as expected, the curves in Fig.~\ref{fig2} (a-c) are not electron-hole symmetric.

Hence, from Fig.~\ref{fig2} we can see that the current on $L$ or $R$ can be switched on or off by changing $\varepsilon_F$, even though the device always remains metallic. The \textit{on} state at the left lead would be achieved when the Fermi energy is on the positive side of the horizontal axis in Fig.~\ref{fig2} (a), while the \textit{off} state would correspond to a Fermi energy on the negative side of the horizontal axis. This is, one gets a switching effect in a metallic device. The magnitude of $\Delta$ needs to allow for an energetic overlap between states from both layers. As $\Delta$ decreases the \textit{on-off} energy region where one way transport can be achieved also does. Taking this into account, a compromise between a satisfactory one way transport energy region and an experimentally affordable $\Delta$ value needs to be achieved. In the following paragraphs we explore the physics behind the directional switching effect in a metal described above, we also numerically prove the valley filtering features of the proposed device and evaluate the robustness of the generated valley currents in the presence of intervalley processes.

To rationalize the non reciprocal electronic transport observed in Fig.~\ref{fig2}, we now turn to the band structure of a partially covered graphene bilayer. Specifically we will show how one can craft asymmetric propagating states under feasible experimental conditions. The four panels of Fig.~\ref{fig3} present the electronic dispersion weighted on the edges of the \textit{bottom layer} of a partially covered “zig-zag” graphene bilayer. The color scale encodes the weight of the corresponding states on a narrow stripe of $6 a$  from the covered ((b) and (d)) or uncovered edge ((a) and (c)).

Panels (a) and (b) correspond to the interlayer interaction set to $\gamma_{1} = 0$, while panels (c) and (d) have $\gamma_{1}$ = 0.15$\gamma_{0}$~\cite{McCann2007}. In these simulations the width ($W$) of the bottom layer is $150a$ and the top layer covers just one fourth of the bottom one ($W/4$).  

Figure~\ref{fig3} shows the energy dispersion of a partially covered bilayer-monolayer junction with (a and b) and without the interlayer interaction. The color scale encodes the weight on a narrow stripe of 6 unit cells from the edges of the bottom layer as represented in the schemes in the insets to each panel. These unveils a peculiar effect of the interlayer interaction: the selective switch-off of the propagating state localized in the covered edge, whose absence is highlighted with a dashed rectangle (panel d). In contrast, the propagating state on the uncovered edge remains as seen in Fig.~\ref{fig3}(c) (red arrows). Interestingly, these surviving states move in one direction without a counterpart propagating on the opposite one, and have a pseudomomentum limited to one valley.

This selective break down of the propagating state localized on the covered edge yields an asymmetrical band structure that is responsible for the non-reciprocal transport properties of the device. At this point one can compare the action of the proposed system with that of a diode. However, in contrast to a diode, our proposal works  even when the bias voltage difference between $L$ and $R$ is zero, and it does not rely on the generation of charge separation in a p-n junction.

\subsection{Valley filtering} 
    
As a final step we turn to the valley filtering features of the device. With this in mind, we turn to a modified 
device in order to quantify the depolarization of the propagating states that make their way to the exit lead (L or R), and to ascertain the persistence of this polarization to valley scatterers. To such end we consider that the magnetic field is slowly turned-off when entering the graphene leads. This provides a unambiguous definition of the valleys in the asymptotic channels. To evaluate the sensitivity of the valley polarization to defects, we randomly place vacancies over a length $250 a$ at each end of the device, immediately after the region where the magnetic field is turned-off. This way, we can test whether the polarization remains after traversing a region with defects and how fast is the depolarization process (note that in this configuration one can have forward intervalley processes).

For this new device scheme the scattering matrix is obtained in the normal modes base which allows to estimate the scattering probabilities connecting the propagating modes of each lead. With this information the degree of polarization of the state at each valley is computed as ${\cal P} = \frac{s(\kappa)-s(\kappa ')}{s(\kappa)+s(\kappa ')}$ where $\kappa$ and $\kappa'$ are the valleys labels and $s(\kappa)$ is the transmission probability summed over all the outgoing modes neighboring $\kappa$ valley. 

\begin{figure}
  \centering
  \includegraphics[width=8cm]{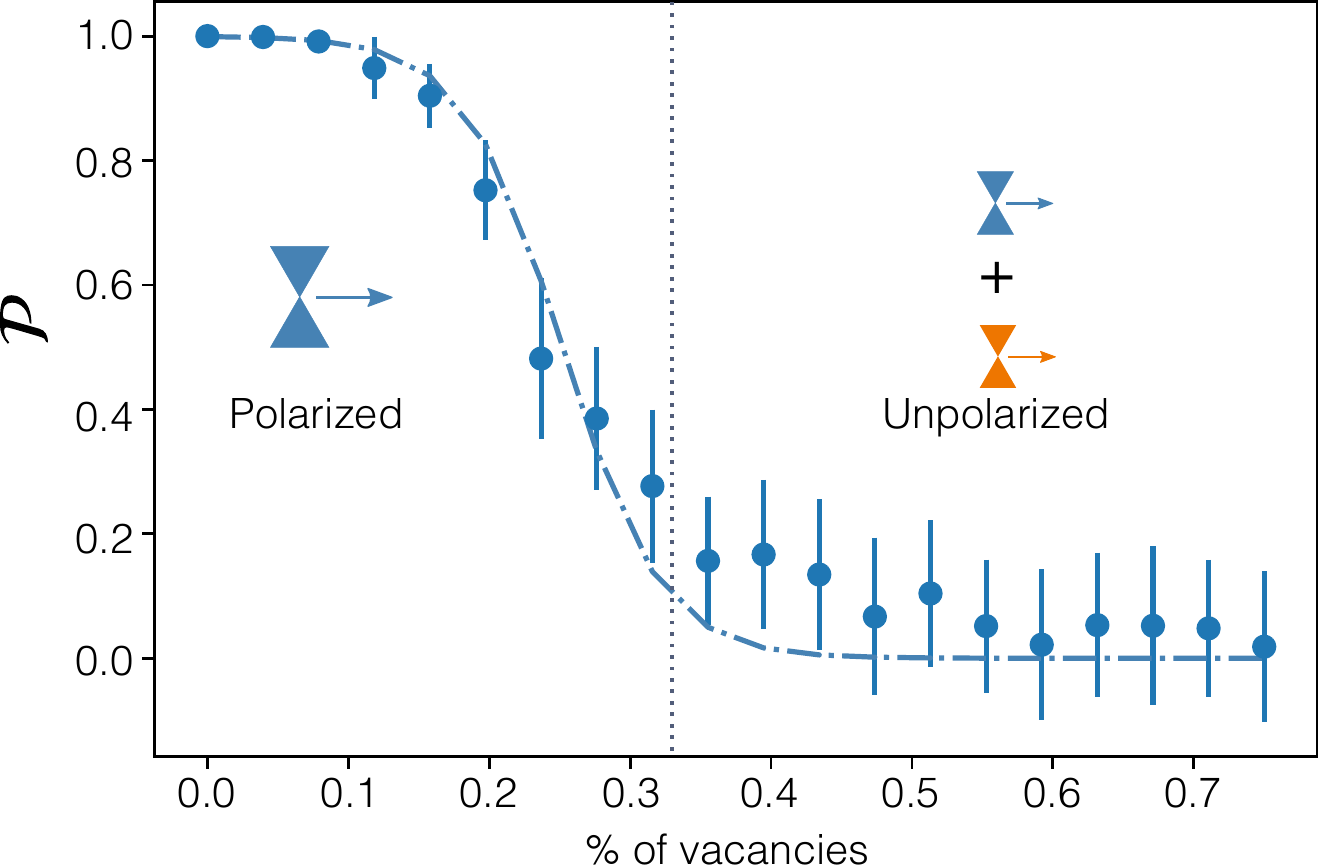}
  \caption{Valley polarization of the outgoing state as a function of the concentration of vacancies. The scatter plot corresponds to the average of $50$ realizations.}
  \label{fig4}
\end{figure}

Figure~\ref{fig4} shows valley polarization, averaged over $50$ realizations, as a function of the vacancies concentration. One can see in absence of vacancies the state that ``survives'' the filtering process is perfectly polarized, as was suggested by the analysis of the band structure. Interestingly, we can note that the polarization of the outcoming states is not immediately annihilated by the presence of valley scatterers, it persists despite the presence of a considerable concentration of vacancies. The fitting of the obtained results (blue dashed line) with an hyperbolic tangent function points out that with $\sim 0.25\%$ of valley vacancies the valley polarization is ${\cal P(\kappa)}=0.5\%$. This suggests that the polarization of the outcoming currents could be experimentally measured and used as a source of valley polarized currents. These sources of switchable valley polarized states remain absent from the toolkit of nano-electronics and is the cornerstone for the expansion of valleytronic devices.

\section{Final remarks} 

We have presented an experimentally feasible way to generate layer-polarized one-way states in an heterogenous bilayer graphene device, a building block for directional \textit{layertronics}~\cite{Jaskolski2018}. The search of a transistor-like effect in a metal has been an outstanding issue for a long time~\cite{Rotkin2004} and one of the motivations for the discovery of graphene~\cite{Novoselov2004} where the modest on-off ratios were considered to be ``a fundamental limitation of any material without a gap exceeding $k_B T$''~\cite{Novoselov2004}. In our case, the on-off ratios are expected to be comparable with state of the art transistors (with the differences pointed out in the introduction).

The proposed system may also serve as a possible source of valley polarized carriers. Interestingly, our results show that the valley polarization may stand a high concentration of defects, thereby opening interesting prospects for further research.

\textit{Author contributions.} All the authors contributed to the planning of the research. MB carried out the numerical simulations and wrote a draft on the main results. LEFFT wrote a first draft of the manuscript which was improved by all authors.

\textit{Conflicts of interest.} There are no conflicts of interest to declare

\textit{Acknowledgments.} MB, JEBV and LEFFT acknowledge support from FondeCyT (Chile) through grants number 3170143, 3170126 and 1170917.

%

\newpage

\begin{figure*}
  \centering
  \includegraphics[width=15cm]{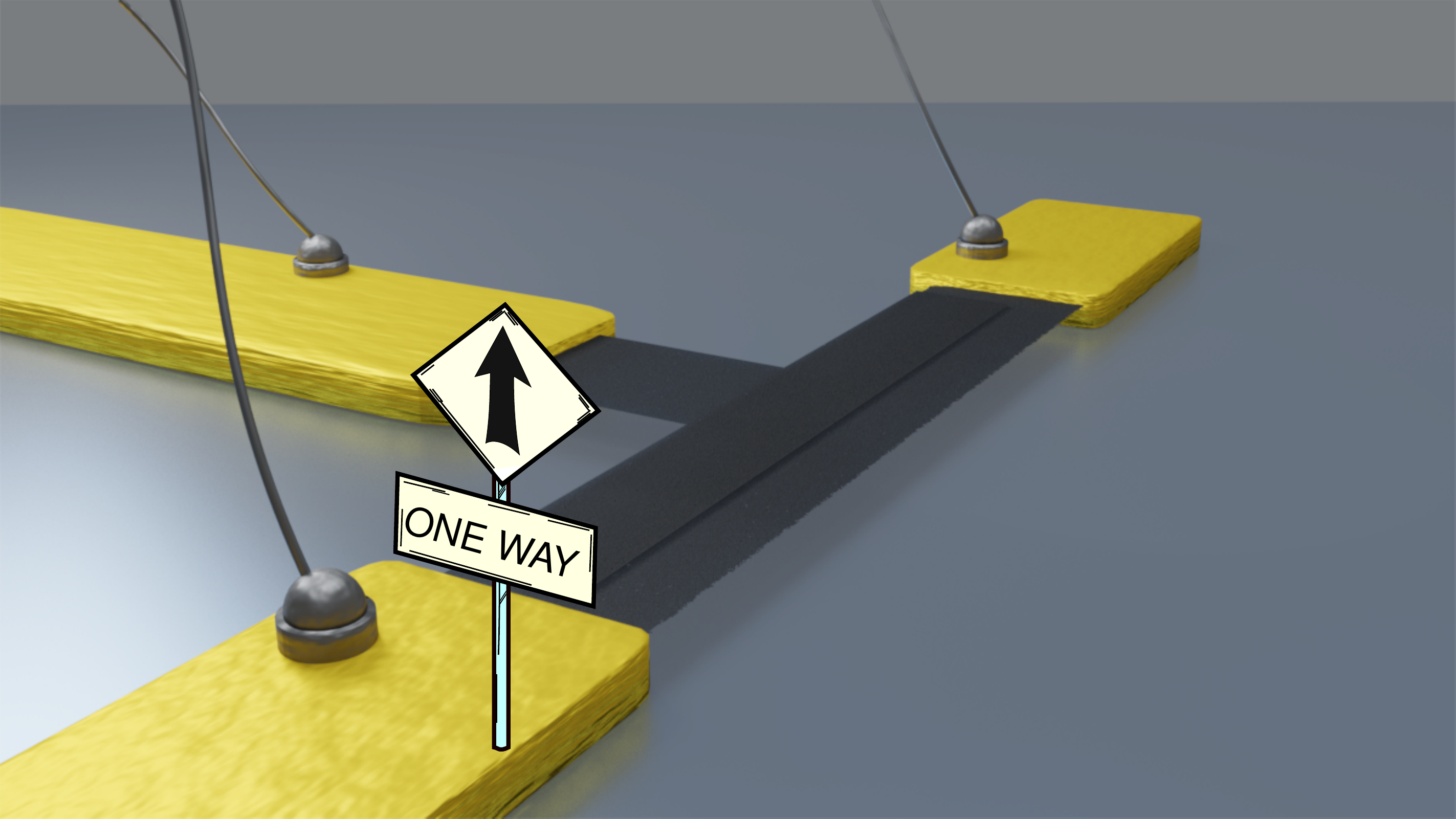}
  \caption{TOC figure.}
  \label{fig5}
\end{figure*}

\end{document}